\documentclass[conference]{IEEEtran}
\IEEEoverridecommandlockouts
\usepackage{cite}
\usepackage{amsmath,amssymb,amsfonts}
\usepackage{graphicx}
\usepackage{textcomp}
\usepackage{xcolor}

\usepackage[utf8]{inputenc}
\usepackage[T5]{fontenc}
\usepackage{url} 
\usepackage[table]{xcolor}
\usepackage[hidelinks]{hyperref}
\usepackage{algorithm}
\usepackage{float}
\usepackage{pdfpages}
\usepackage{svg} 
\usepackage{amsmath}
\usepackage{booktabs}
\usepackage{tabularx}
\usepackage{algpseudocode}
\usepackage{stfloats}
\usepackage{tabularx}
\usepackage{booktabs}
\usepackage{orcidlink}
\usepackage{authblk}

\usepackage{caption} 
\def\BibTeX{{\rm B\kern-.05em{\sc i\kern-.025em b}\kern-.08em
    T\kern-.1667em\lower.7ex\hbox{E}\kern-.125emX}}

\begin{document}

\title{ViFA-Council: Multi-Agent LLM Deliberation for Vietnamese Folk Art Generation}

\author[1,2]{Hai-Dang Nguyen\orcidlink{0009-0002-6475-6299}}
\author[1,2]{Minh-Phuong Pham\orcidlink{0009-0000-8063-7762
}}
\author[1,2]{Thao Thi Phuong Dao\orcidlink{0000-0002-0109-1114}}
\author[1,2]{Trong-Le Do\orcidlink{0000-0002-2906-0360}}
\author[2,3]{\\Vinh-Tiep Nguyen\orcidlink{0000-0003-4260-7874}}
\author[1,2]{Trung-Nghia Le \orcidlink{0000-0002-7363-2610}$^\dagger$\thanks{$^\dagger$Corresponding author: ltnghia@fit.hcmus.edu.vn}}

\affil[1]{University of Science, Ho Chi Minh City, Vietnam}
\affil[2]{Vietnam National University, Ho Chi Minh City, Vietnam}
\affil[3]{University of Information Technology, Ho Chi Minh City, Vietnam}

\maketitle

\begin{abstract}
This paper presents ViFA-Council, a three-stage multi-agent framework that employs multiple large language models (LLMs) to tackle two culturally complex generative tasks: image outpainting and educational story generation based on traditional Vietnamese folk paintings. Current single-model generative pipelines frequently struggle with stylistic hallucinations and cultural misrepresentations because they lack mechanisms for cross-model critique. ViFA-Council addresses this challenge by orchestrating collaboration among GPT-4o, Gemini 3.1 Pro, and Claude Sonnet 4.6. It enforces rigorous cultural constraints through structured agent deliberation. This deliberation is mediated by task-specific JSON schemas that effectively bridge natural language discussions with diffusion-based image synthesis using Banana Pro. Experiments and a user study demonstrate that structured multi-agent deliberation is a promising direction for improving cultural fidelity and narrative coherence in culturally sensitive, low-resource artistic domains. The source code and data are released at \url{https://github.com/DanielNguyen-05/ViFA-Council}.
\end{abstract}


\section{Introduction}

Large language models (LLMs) have achieved remarkable success in multimodal generation. However, generating culturally faithful content for low-resource artistic traditions remains an open challenge. Single-model pipelines frequently produce stylistic hallucinations and cultural misrepresentations, as individual models have no mechanism to cross-check or critique each other's outputs~\cite{cao2023comprehensive}.

Traditional Vietnamese folk paintings represent a centuries-old tradition conveying communal values, auspicious symbolism, and narratives drawn from rural life and spirituality. Yet in recent decades, the number of skilled artisans has declined sharply due to limited interest from younger generations and shrinking demand for woodblock prints. Recognizing the urgency of this loss, UNESCO has inscribed Đông Hồ folk painting (Bắc Ninh) on the List of Intangible Cultural Heritage in Need of Urgent Safeguarding~\cite{UNESCO}, and the Vietnamese government has since launched dedicated preservation programs to sustain both the production and transmission of this heritage. Despite this institutional recognition, preservation efforts must go beyond archiving: they should be engaging, accessible, and pedagogically rich. Paintings such as \textit{Thầy Đồ Cóc} encode moral lessons through symbolic personification ideal for children's education, yet virtually no AI-based tools exist to transform folk paintings into modern, engaging storytelling formats while maintaining their authentic artistic character.

\begin{figure}[t!]
\centering
\includegraphics[width=\linewidth]{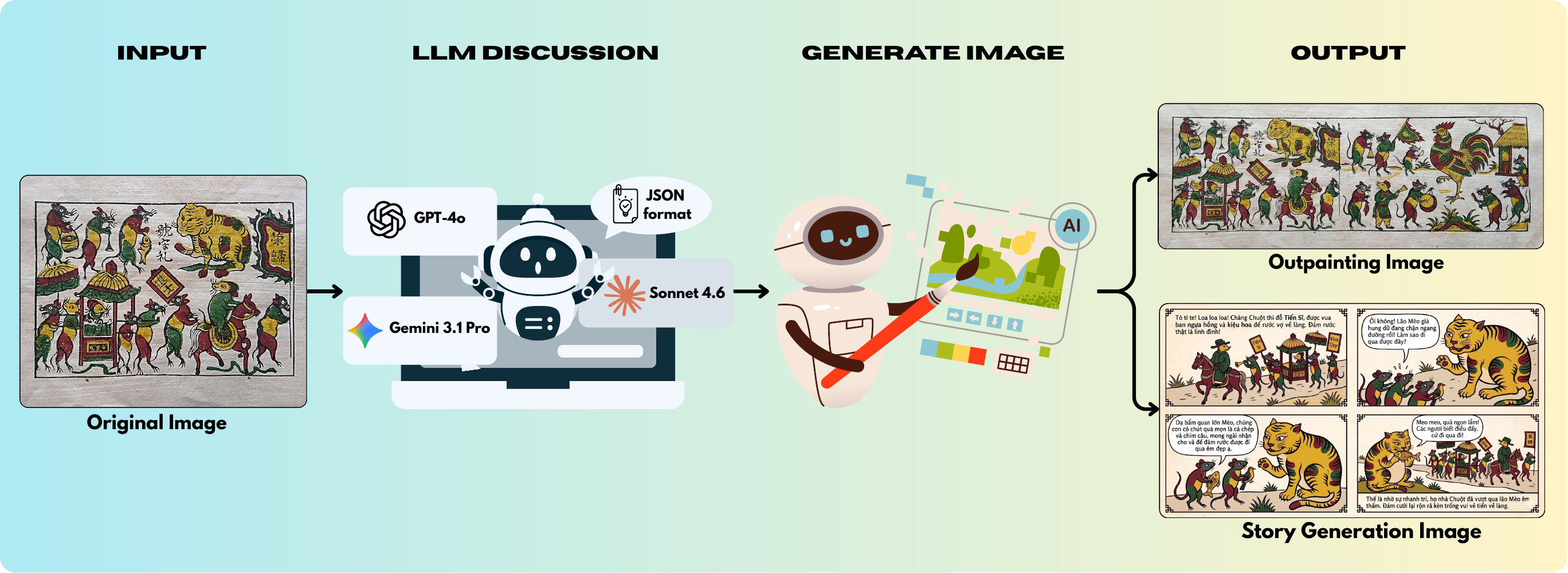}
\caption{Overview of the ViFA-Council multi-agent framework, which actively transforms traditional Vietnamese folk paintings through complex generative tasks including image outpainting and the creation of educational stories.}
\label{fig:teaser}
\vspace{-5mm}
\end{figure}

To address this gap, we propose \textbf{ViFA-Council}, a robust three-stage multi-agent framework that seamlessly integrates structured deliberation with culturally enforced generative constraints (Fig.~\ref{fig:teaser}). The system operates through parallel drafting in Stage~1, rigorous cross-model refinement in Stage~2, and Chairman-based adjudication in Stage~3. Every stage is governed by task-specific JSON schemas that act as a bridge between the complex language model discourse and downstream image synthesis. To the best of our knowledge, ViFA-Council is the first framework applying multi-agent LLM deliberation to Vietnamese folk art preservation and AI-assisted folk art education.

\begin{figure*}[t!]
\centering
\includegraphics[width=0.95\linewidth]{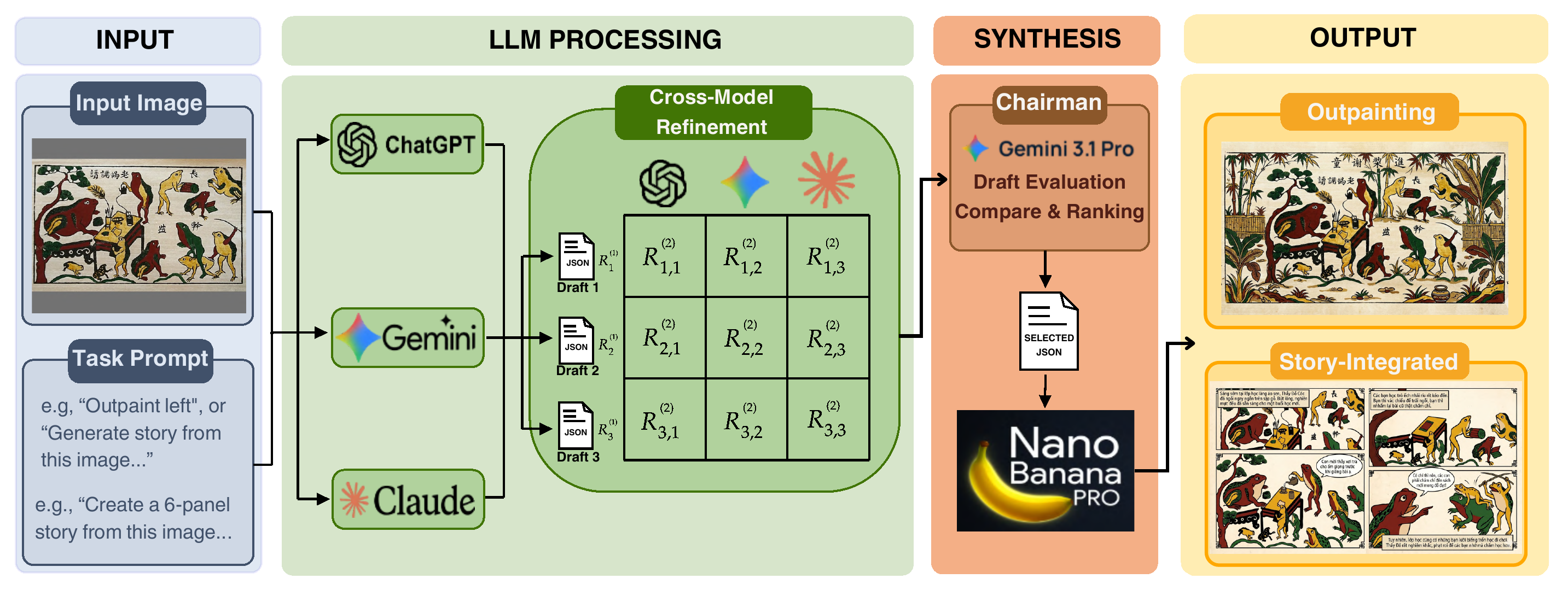}
\vspace{-2mm}
\caption{Detailed visual breakdown of the proposed three-stage LLM Council architecture designed for Vietnamese folk art generation. This pipeline highlights the sequential progression from initial parallel drafting, through rigorous cross-model refinement, and concluding with final Chairman-based adjudication.}
\label{fig:workflow}
\vspace{-3mm}
\end{figure*}

We evaluated ViFA-Council on a curated collection of over 200 traditional Vietnamese folk paintings from the \textit{Đông Hồ} (Bắc Ninh), \textit{Hàng Trống} (Hà Nội), and \textit{Kim Hoàng} (Hà Tây) traditions, sourced from publicly accessible museum websites, cultural heritage repositories, and reputable news outlets. 
Compared to standalone GPT-4o, Gemini 3.1 Pro, and Claude Sonnet 4.6 baselines, our system recorded the highest scores on the automated boundary-blending metrics for outpainting and on narrative-coherence metrics for story generation. Additionally, a user study confirmed these findings, with human raters heavily favoring ViFA-Council on the dimensions of cultural relevance and dialogue quality. The source code and data are released at \url{https://github.com/DanielNguyen-05/ViFA-Council}.

The main contributions of this paper are as follows:
\begin{itemize}
    \item We introduce a multi-agent deliberation system specifically to the AI-assisted preservation and educational adaptation of Vietnamese folk art.
    \item We propose a structured, three-stage Council that mitigates cultural and stylistic hallucinations through cross-model critique and adjudication.
    \item We design task-specific JSON schemas that translate high-level cultural constraints into structured and parameterized image-generation instructions.
\end{itemize}

\section{Related Work}

\subsection{Multi-Agent LLM Systems}

Multi-agent debate has been shown to significantly improve factual 
accuracy and reasoning over single-model inference~\cite{du2024improving}. ChatEval~\cite{chan2023chateval} further demonstrated that using models with diverse roles leads to better evaluation quality than homogeneous ensembles. Related work has explored role specialization, communicating agent societies, and structured debate for creative tasks~\cite{liang2024encouraging}. However, these systems primarily focus on text-based reasoning. Our work extends this paradigm by introducing a structured JSON interface that bridges LLM deliberation with downstream image synthesis, and an $M \times N$ cross-refinement protocol that separates authorship from critique more rigorously than round-robin approaches, a property especially critical when model biases are systematic rather than random.

\subsection{Image Outpainting and Cultural Heritage AI}

Latent Diffusion Models~\cite{rombach2022high} now underpin most inpainting and outpainting pipelines, with LLM-guided methods such as Yang et al.~\cite{yang2024idea2img} extending control to language-driven expansion. However, these approaches target photographs or generic art and cannot enforce the strict iconographic conventions of folk traditions. On the preservation side, generative AI has been applied to East Asian heritage such as Cantonese porcelain and Chinese New Year prints, while Vietnamese folk art remains limited to scanning and archiving. The closest related work, Nguyen et al.~\cite{nguyen2025can}, applies LLMs to the Vietnamese board game Ô Ăn Quan but focuses on decision-making rather than creative generation and uses no deliberation mechanism. To our knowledge, ViFA-Council is the first system to combine multi-agent deliberation with structured JSON-to-image synthesis for Vietnamese intangible cultural heritage.

\subsection{LLMs for Structured Visual Generation}

Using LLMs to produce structured specifications, rather than pixels directly, has emerged as an effective strategy for controllable image synthesis. Chain-of-thought prompting and persona-based agent modeling~\cite{salemi2024lamp} improve both output quality and role consistency. Our prompt design builds on these ideas: Stage~1 agents reason about cultural context before completing the JSON schema, and the Chairman (Stage~3) must justify its selection before returning a decision.

\section{Proposed Method}

\subsection{Overview}

We propose a unified multi-agent deliberation framework for folk-painting outpainting and educational story generation, through a structured three-stage collaboration. As shown in Fig.~\ref{fig:workflow}, the system takes a Vietnamese folk painting as input and produces a task-specific JSON configuration that drives downstream image synthesis, progressively strengthening stylistic and cultural constraints before any image is generated.

\begin{table*}[t]
\centering
\caption{Comprehensive outline of the essential JSON schema parameters specifically engineered to bridge language model deliberation and visual synthesis for both the image outpainting and educational story generation tasks.}
\label{tab:schema}
\small
\setlength{\tabcolsep}{6pt}
\begin{tabular}{llp{5.5cm}p{4.5cm}}
\toprule
\textbf{Task} & \textbf{Field} & \textbf{Description} & \textbf{Purpose} \\
\midrule
Outpainting & \texttt{input\_image\_analysis}
  & Original style and dominant colors & Source-aware visual analysis \\[3pt]
  Outpainting & \texttt{expansion\_settings}
  & Direction, pixel amount, and mask blur & Synthesis spatial control \\[3pt]
Outpainting & \texttt{seamless\_blending}
  & Style, texture, color, pattern constraints & Boundary \& appearance continuity \\[3pt]
Outpainting & \texttt{outpainting\_scenarios[]}
  & Expansion proposals with prompts 
  & Multi-candidate generation \\[3pt]
\midrule
Story Gen.  & \texttt{source\_material}
  & Title, era, cultural context, and symbols
  & Cultural grounding \\[3pt]
Story Gen.  & \texttt{art\_style\_preservation}
  & Character, line, and perspective constraints
  & Prevent visual-style drift \\[3pt]
Story Gen.  & \texttt{characters[]}
  & Appearance, style, and cultural significance
  & Cross-panel character consistency \\[3pt]
Story Gen.  & \texttt{story\_panels[]}
  & Layout, dialogue, motifs, and transitions
  & Sequential narrative specification \\[3pt]
Story Gen.  & \texttt{educational\_content}
  & Objectives, explanations, and glossary
  & Educational adaptation \\
\bottomrule
\end{tabular}
\vspace{-3mm}
\end{table*}

\subsection{LLM Council}

\subsubsection{Task Formulation}

Given a folk painting image $I$ and a user directive $q$, the Council produces a Chairman-selected JSON configuration:
\begin{equation}
J^* = \text{Council}(I, q, \{M_1, \dots, M_N\}).
\end{equation}
\noindent where $M_i$ denotes the $i$-th participating LLM. As in Alg.~\ref{alg:council} , the Council operates through three stages using task-specific prompt templates: $P_{\mathrm{task}}$ for initial draft generation, $P_{\mathrm{refine}}$ for cross-model refinement, and $P_{\mathrm{eval}}$ for Chairman evaluation.

Two task-specific schemas are defined: an \textit{Outpainting Schema} encoding spatial expansion parameters and generation prompts, and a \textit{Story Generation Schema} encoding per-panel layout, character descriptions, dialogue, and style constraints. 

\subsubsection{Stage~1: Parallel Draft Generation}

Each Council member independently generates a candidate JSON from the same input $(I, q)$ without observing the others' outputs, ensuring diverse model priors:
\begin{equation}
R_i^{(1)} = \text{LLM}_i(I, q, P_{\text{task}}), \qquad i = 1, \ldots, N
\end{equation}
All calls are dispatched concurrently with exponential backoff retries (up to three attempts). In our implementation, $N = 3$ agents (GPT-4o, Gemini 3.1 Pro, Claude Sonnet 4.6) produce three independent drafts.

\subsubsection{Stage~2: Cross-Model Refinement}

Stage~2 implements a \textit{cross-refinement protocol} in which every 
agent reviews and improves every Stage~1 draft regardless of authorship, 
producing $M \times N$ refined candidates:
\begin{equation}
R_{i,j}^{(2)} = \text{LLM}_j(I, R_i^{(1)}, P_{\text{refine}}), 
\qquad i=1..M,\; j=1..N
\end{equation}
$P_{\text{refine}}$ instructs each model to verify structural validity, improve cultural fidelity, and return an enhanced JSON in the same schema. All $M \times N$ tasks run in parallel; failed calls fall back to the original Stage~1 draft to ensure no candidate is lost. With $M = N = 3$, Stage~2 produces up to 9 refined candidates per input.

% \begin{figure}[t!]
% \centering
% \includegraphics[width=\linewidth]{img/stage2.pdf}
% \caption{Stage~2 cross-model refinement matrix ($3\times3$). Each cell 
% $R_{i,j}$ denotes the output of agent $j$ refining Stage~1 draft $i$.}
% \label{fig:stage2}
% \end{figure}

\subsubsection{Stage~3: Chairman Evaluation and Selection}

A dedicated Chairman, a separate Gemini 3.1 Pro instance isolated from all prior steps, receives all Stage~1 and Stage~2 candidates as a consolidated prompt, labeled alphabetically to prevent positional bias. The Chairman evaluates each candidate on four criteria: (i) JSON structural validity, (ii) logical coherence of expansion or narrative parameters, (iii) cultural fidelity to Vietnamese folk art conventions, and (iv) whether Stage~2 refinements improve the Stage~1 version. The optimal configuration is selected as:
\begin{equation}
J^{*} = \text{Candidates}\left[\arg\max_{k} \; \text{Score}(C_k)\right].
\end{equation}

The Chairman emits its decision as BEST RESPONSE: Response X, parsed via regular expression; on failure, the system falls back to the top-ranked Stage~2 output.

\subsection{JSON Schema Design and Prompt Engineering}

A central contribution of this work is the design of task-specific JSON schemas that provide a structured, model-independent bridge between LLM deliberation and downstream image generation. Table~\ref{tab:schema} presents the main schema fields.

\begin{algorithm}[t]
\caption{LLM Council Execution}
\label{alg:council}
\small
\begin{algorithmic}[1]
\Require Painting $I$, directive $q$, agents 
  $\{\mathrm{LLM}_1,\dots,\mathrm{LLM}_N\}$
\Ensure Optimal JSON configuration $J^{*}$
\Statex \textit{// Stage~1: Parallel Drafting}
\For{$i = 1 \dots N$ \textbf{in parallel}}
  \State $R_i^{(1)} \leftarrow \mathrm{LLM}_i(I, q, P_{\mathrm{task}})$
\EndFor
\Statex \textit{// Stage~2: Cross-Model Refinement}
\For{$i = 1 \dots M,\; j = 1 \dots N$ \textbf{in parallel}}
  \State $R_{i,j}^{(2)} \leftarrow 
    \mathrm{LLM}_j(I, R_i^{(1)}, P_{\mathrm{refine}})$
\EndFor
\Statex \textit{// Stage~3: Chairman Selection}
\State Candidates $\leftarrow \{R_i^{(1)}\} \cup \{R_{i,j}^{(2)}\}$
\State $J^{*} \leftarrow \mathrm{Chairman}(\text{Candidates}, P_{\mathrm{eval}})$
\State \Return $J^{*}$
\end{algorithmic}
\end{algorithm}

The \textit{Outpainting Schema} includes \texttt{input\_image\_analysis} for identifying the source style and dominant colors, \texttt{expansion\_settings} for controlling spatial expansion, and \texttt{seamless\_blending} for preserving texture, color, and pattern continuity. It also contains \texttt{outpainting\_scenarios[]}, which specifies alternative expansion descriptions and synthesis prompts.

The \textit{Story Generation Schema} uses \texttt{source\_material} to encode the painting's cultural context and symbolic elements, while \texttt{art\_style\_preservation} defines constraints on character appearance, palette, line quality, and perspective. The \texttt{characters[]} and \texttt{story\_panels} fields maintain character consistency and organize the sequential narrative, whereas \texttt{educational\_content} provides learning objectives, cultural explanations, and glossary entries. By separating style-preservation constraints from generative content, the schemas help reduce visual drift toward modern or generic aesthetics.

\subsection{Image Synthesis via Banana Pro}

The selected $J^{*}$ is passed together with the input image $I$ to Banana Pro model. For outpainting, the \texttt{expansion\_settings} field specifies the expansion region, while \texttt{seamless\_blending} provides style, texture, color, and pattern-continuity constraints. The selected \texttt{outpainting\_scenarios[]} entry supplies the final synthesis prompt. For story generation, each \texttt{story\_panels} specification is rendered sequentially, with visual constraints enforced by \texttt{art\_style\_preservation} to preserve character appearance, color palette, line quality, and perspective throughout the story. This structured JSON-to-inference mapping keeps the synthesis layer model-agnostic, allowing compatible diffusion backends to be substituted without modifying the upstream Council framework.

\section{Experiments}

\begin{figure*}[t]
\centering
\small
\begin{tabular}{ccccc}
\makebox[0.13\linewidth][l]{\textbf{Original}} &
\makebox[0.33\linewidth][c]{\textbf{ViFA-Council (Ours)}} &
\makebox[0.135\linewidth][c]{\textbf{GPT-4o}} &
\makebox[0.145\linewidth][c]{\textbf{Gemini 3.1 Pro}} &
\makebox[0.15\linewidth][c]{\textbf{Claude Sonnet 4.6}}
\vspace{0.1cm}
\end{tabular}
\includegraphics[width=\linewidth]{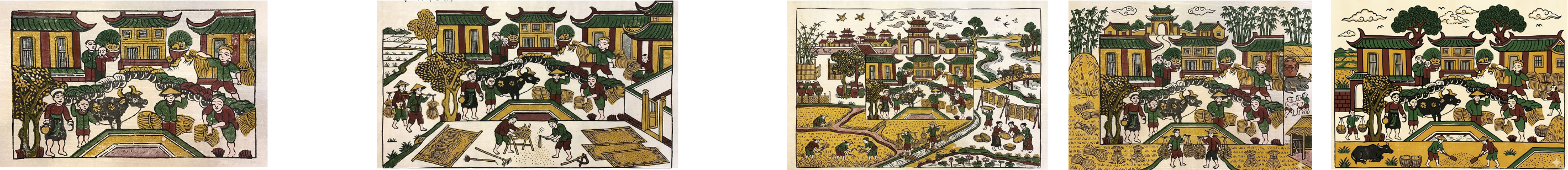}\\ \vspace{0.25cm}
\includegraphics[width=\linewidth]{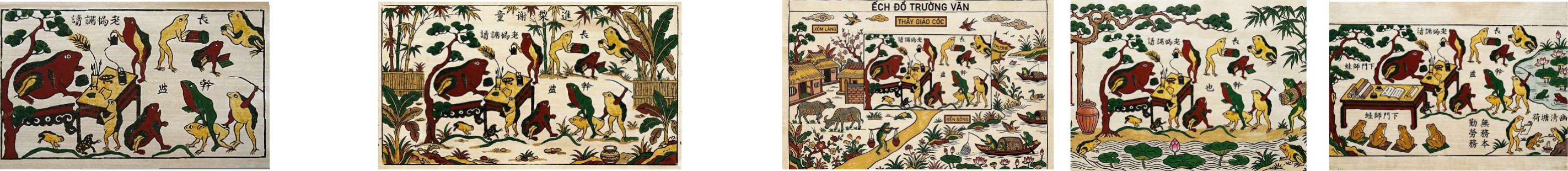}\\ 
\vspace{0.25cm}
\includegraphics[width=\linewidth]{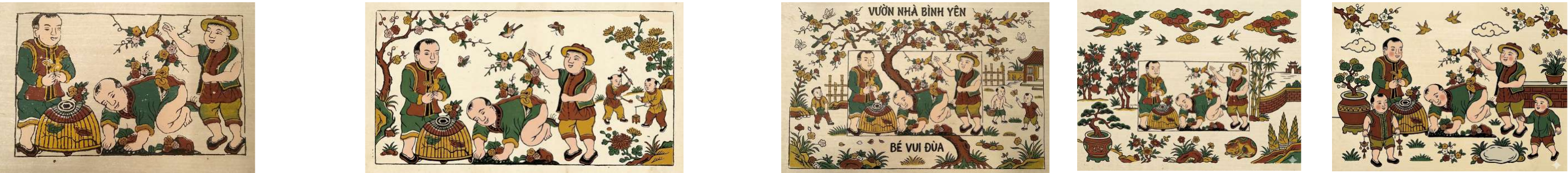}\\ \vspace{0.25cm}
\includegraphics[width=\linewidth]{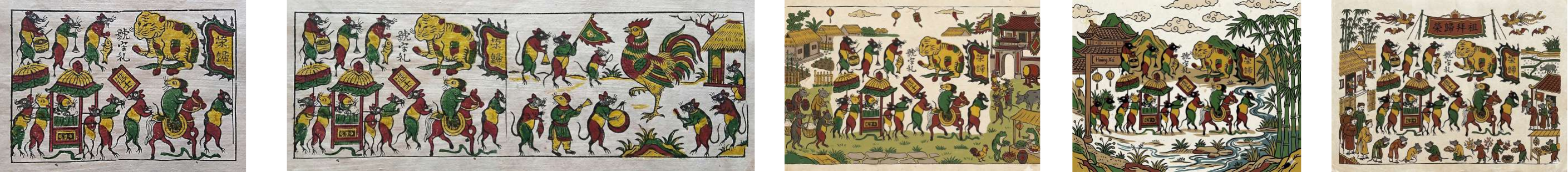}\\
\caption{Visual comparison of outpainting results. ViFA seamlessly extends the narrative while preserving Vietnamese folk-art-inspired textures, whereas baselines exhibit cultural hallucinations and textural degradation.}
\label{fig:outpainting_comparison}
\vspace{-3mm}
\end{figure*}

\begin{figure*}[t]
\centering
\small
\begin{tabular}{ccc}
\makebox[0.2\linewidth][l]{\textbf{Original}} &
\makebox[0.5\linewidth][c]{\textbf{ViFA-Council (Ours)}} &
\makebox[0.16\linewidth][c]{\textbf{Gemini 3.1 Pro}}
\vspace{0.1cm}
\end{tabular}
\includegraphics[width=\linewidth]{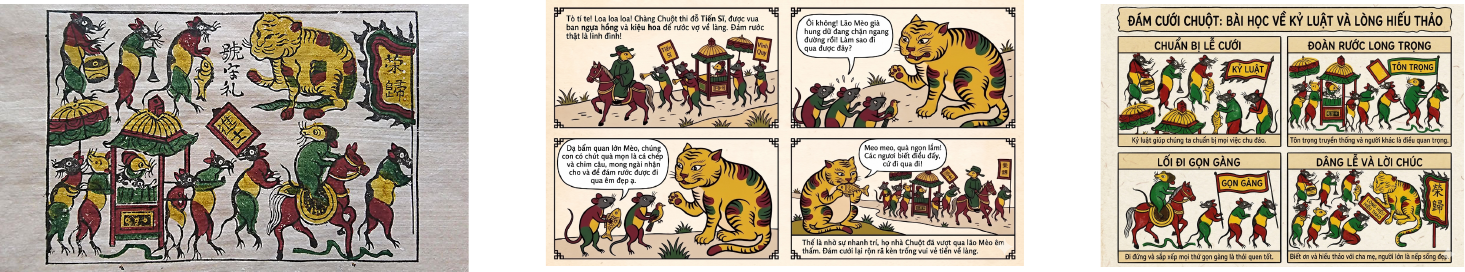}\\ \vspace{0.25cm}
\includegraphics[width=\linewidth]{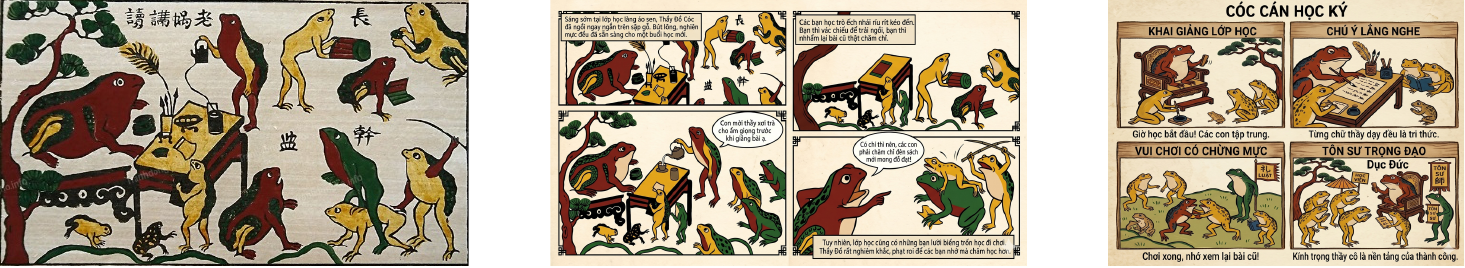}\\ \vspace{0.25cm}
\includegraphics[width=\linewidth]{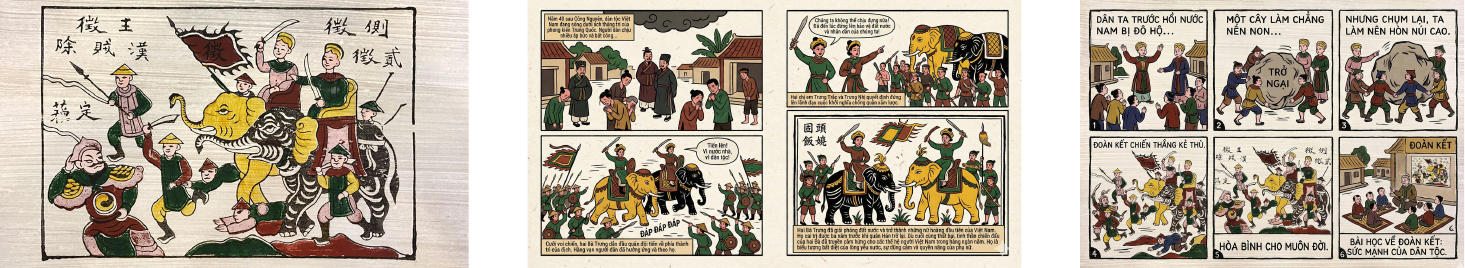}\\
\caption{Visual comparison of story generation results. ViFA preserves the Vietnamese folk-art comic-strip format with culturally grounded dialogue, whereas Gemini produces modern infographic-style layouts inconsistent with the Đông Hồ tradition.}
\label{fig:story_comparison}
\vspace{-3mm}
\end{figure*}

\subsection{Experimental Setup}
\label{sec:setup}

All experiments are conducted on a curated collection of 257 Vietnamese folk paintings, comprising 112 \textit{Đông Hồ}, 74 \textit{Hàng Trống}, and 71 \textit{Kim Hoàng} works, annotated with school origin, iconographic subject category, and production era. The Council coordinates GPT-4o, Gemini 3.1 Pro, and Claude Sonnet 4.6 as Expert Agents, with a dedicated Gemini 3.1 Pro instance serving as Chairman. All API calls use temperature $0.7$, dispatched concurrently via \texttt{asyncio.gather()} with exponential backoff. Image synthesis is performed via Banana Pro, using diffusion-based outpainting for Task 1 and sequential text-to-image generation for Task 2.

For \textbf{Task~1 (Outpainting)}, we report \textit{Style Consistency}, using CLIP~\cite{radford2021learning} similarity between the generated region and reference paintings from the corresponding folk-art tradition, together with \textit{Content Consistency} measured by LPIPS~\cite{zhang2018unreasonable} and DreamSim~\cite{fu2023dreamsim} to assess perceptual and structural blending quality. For \textbf{Task~2 (Story Generation)}, we report DINOv2~\cite{oquab2023dinov2} to measure protagonist appearance consistency across panels, together with PickScore~\cite{kirstain2023pick} for text-image alignment. Across both tasks, we additionally report a \textit{JSON Semantic Score} (0--100), which evaluates schema completeness, task coherence, culture-specific grounding, and the absence of foreign or modern stylistic elements.

\subsection{Qualitative Evaluation}
\label{sec:qualitative}

Single-model baselines exhibit four recurring failure modes across both tasks. \textbf{Typographical Content Injection}: GPT-4o hallucinates modern Vietnamese text directly onto the canvas (e.g., ``VƯỜN NHÀ BÌNH YÊN'', ``BÉ VUI ĐÙA''), violating the aniconic conventions of folk painting (Fig.~\ref{fig:outpainting_comparison}, rows 2--3). \textbf{Cultural Hallucination}: Gemini injects Sino-centric motifs, imperial swirling clouds, ornamental rock gardens, and curved-roof pagodas, that replace naive rural aesthetics with generic East Asian tropes (Fig.~\ref{fig:outpainting_comparison}, rows 3--4). \textbf{Textural and Chromatic Degradation}: Claude produces structurally plausible boundaries but flattens the earthy palette and loses the fibrous texture of \textit{giấy điệp} (seashell-coated bark paper), yielding a sanitized digital appearance. \textbf{Structural Format Drift}: In the story generation task, Gemini morphs comic panels into modern educational infographics with rigid typographic headers (e.g.,``ĐÁM CƯỚI CHUỘT: BÀI HỌC VỀ KỶ LUẬT'', ``KHAI GIẢNG LỚP HỌC''), obliterating the Đông Hồ aesthetic entirely (Fig.~\ref{fig:story_comparison}).

In contrast, ViFA-Council systematically avoids all four pitfalls. For outpainting, it achieves seamless boundary blending while anchoring extensions to authentic folk colors, textures, and motifs (Fig.~\ref{fig:outpainting_comparison}). For story generation, it preserves the traditional multi-panel layout with organic speech bubbles and strict character consistency across complex anthropomorphic designs (Fig.~\ref{fig:story_comparison}).

\subsection{Comparison against Single-Model Baselines}

We compare ViFA-Council against three single-model baselines, GPT-4o, Gemini 3.1 Pro, and Claude Sonnet 4.6, each prompted with the identical Stage~1 instruction without cross-refinement or Chairman adjudication, using the same Banana Pro pipeline for a fair comparison.

\begin{table}[t!]
\centering
\caption{Quantitative experimental results for the image outpainting task. DS and JSS denote DreamSim and JSON Semantic. The best results are shown in \textbf{bold}.}
\label{tab:outpainting}
% \resizebox{0.48\textwidth}{!}{
\begin{tabular}{lcccc}
\toprule
\textbf{Model} & \textbf{CLIP$\uparrow$} & \textbf{LPIPS$\downarrow$} 
& \textbf{DS $\downarrow$} & \textbf{JSS (\%)\,$\uparrow$} \\
\midrule
ChatGPT                      & 0.8348 & 0.7832 & 0.5896 & 70.10 \\
Claude                       & 0.8485 & 0.7986 & 0.6169 & 84.30 \\
Gemini                       & 0.8520 & 0.7705 & 0.6134 & 76.70 \\
\textbf{ViFA-Council (Ours)} & \textbf{0.8524} & \textbf{0.7317} 
                             & \textbf{0.5745} & \textbf{93.90} \\
\bottomrule
\end{tabular}
% }
\vspace{-3mm}
\end{table}

\begin{table}[t!]
\centering
\caption{Quantitative comparison between ViFA-Council and the selected Gemini 3.1 Pro single-model baseline on educational story generation. PS and JSS denote PickScore and JSON Semantic. The best results are shown in \textbf{bold}.}
\label{tab:story}
% \resizebox{0.48\textwidth}{!}{
\begin{tabular}{lcc}
\begin{tabular}{lcccc}
\toprule
\textbf{Model} & \textbf{CLIP $\uparrow$} & \textbf{DINO $\uparrow$} & \textbf{PS $\uparrow$} & \textbf{JSS (\%) $\uparrow$} \\
\midrule
Gemini 3.1 Pro & 0.667 & 0.504 & 18.573 & 87.0 \\
\textbf{ViFA-Council (Ours)} & \textbf{0.703} & \textbf{0.601} & \textbf{19.61} & \textbf{93.0} \\
\bottomrule
\end{tabular}
\end{tabular}
% }
\vspace{-3mm}
\end{table}

ViFA-Council achieves the strongest overall quantitative performance across both tasks. In outpainting, it delivers the best boundary blending and highest stylistic fidelity (Table~\ref{tab:outpainting}). In story generation, it outperforms Gemini in Narrative Coherence and Text-to-Image Alignment (Table~\ref{tab:story}). The largest gain is in the JSON Semantic Score, improving by $9.6\%$ over Claude for outpainting and $6.0\%$ over Gemini for story generation, indicating that the Council pipeline better enforces cultural constraints than single-model inference. Notably, Gemini attains competitive visual metrics despite hallucinating Sino-centric motifs, demonstrating that pixel-level metrics alone cannot capture cultural fidelity and reinforcing the need for the human evaluation in Section~\ref{sec:userstudy}.

\subsection{User Study}
\label{sec:userstudy}

To complement the automated evaluation, we conducted a blind user study with 15 participants (10 STEM students, 3 educators, and 2 AI engineers), all familiar with LLM-based generative AI. Model identities were anonymized, output order was randomized, and participants rated each result on a five-point Likert scale  (1 = Strongly Disagree, 5 = Strongly Agree) across task-specific dimensions.

As shown in Tables~\ref{tab:user_outpainting} and~\ref{tab:user_storygen}, ViFA-Council receives the highest overall ratings for both tasks. For outpainting, participants consistently preferred its seamless blending, visual consistency, and perceived cultural relevance over the single-model baselines. For story generation, ViFA-Council achieves the largest improvements in Dialogue Quality ($3.96$ vs.\ $3.54$) and Narrative Coherence ($3.94$ vs.\ $3.66$), while avoiding the modern infographic layouts frequently observed in the Gemini baseline. The lower rating variance also indicates greater agreement among participants regarding the quality of ViFA-Council outputs.

\begin{table}[t!]
\centering
\caption{Human evaluation ratings for the outpainting task based on a five-point Likert scale. SB, SC, CR, and OS denote Seamless Blending, Style Consistency, Cultural Relevance, and Overall Score. The best results are shown in \textbf{bold}.}
\label{tab:user_outpainting}
% \resizebox{0.48\textwidth}{!}{
\begin{tabular}{lcccc}
\toprule
\textbf{Method} & \textbf{SB$\uparrow$} & \textbf{SC$\uparrow$} 
& \textbf{CR$\uparrow$} & \textbf{OS$\uparrow$} \\
\midrule
GPT-4o                       & 3.18 & 3.42 & 3.52 & 3.37 $\pm$ 1.16 \\
Gemini 3.1 Pro               & 3.36 & 3.48 & 3.55 & 3.46 $\pm$ 1.11 \\
Claude Sonnet 4.6            & 3.95 & 3.64 & 3.96 & 3.85 $\pm$ 1.00 \\
\textbf{ViFA-Council (Ours)} & \textbf{4.08} & \textbf{4.08} 
                             & \textbf{4.00} & \textbf{4.05 $\pm$ 0.89} \\
\bottomrule
\end{tabular}
% }
\end{table}

\begin{table}[t!]
\centering
\caption{Human evaluation ratings for the story generation task based on a five-point Likert scale. CR, NC, DQ, and OS denote Cultural Relevance, Narrative Coherence, Dialogue Quality, and Overall Score. The best results are shown in \textbf{bold}.}
\label{tab:user_storygen}
\begin{tabular}{lcccc}
\toprule
\textbf{Method} & \textbf{CR$\uparrow$} & \textbf{NC$\uparrow$} & \textbf{DQ$\uparrow$} & \textbf{OS$\uparrow$} \\
\midrule
Gemini 3.1 Pro & 3.56 & 3.66 & 3.54 & 3.59 $\pm$ 1.05 \\
\textbf{ViFA-Council (Ours)} & \textbf{3.89} & \textbf{3.94} & \textbf{3.96} & \textbf{3.93 $\pm$ 0.88} \\
\bottomrule
\end{tabular}
% }
\vspace{-3mm}
\end{table}

\subsection{Limitations}

ViFA-Council is evaluated in a zero-shot setting without fine-tuning on Vietnamese folk-art datasets, which may limit adaptation to subtle school-specific visual conventions. Although the user study measures perceived cultural relevance, it does not include professional folk-art historians or artisans; therefore, historical and iconographic authenticity remains to be validated through expert assessment. As an orchestration framework, ViFA-Council remains dependent on the capabilities of the underlying LLMs and image-generation models, and limitations in these foundation models may still propagate to the final outputs. In addition, the multi-agent pipeline requires multiple parallel API calls, increasing inference cost and latency compared with single-model generation.

\section{Conclusion}
\label{sec:conclusion}

We presented ViFA-Council, a three-stage multi-agent framework for culturally sensitive generative tasks such as folk-painting outpainting and educational story generation. The complete Council pipeline helps reduce several common single-model failure modes, including cultural hallucination and structural format drift. Through task-specific JSON schemas, the framework provides a structured, model-agnostic bridge between language-model reasoning and downstream visual synthesis. Beyond its technical contribution, ViFA-Council supports the transformation of Vietnamese folk paintings into accessible comic-style narratives for educational use. These findings highlight the potential of multi-agent generative systems for culturally informed digital engagement and the educational adaptation of low-resource cultural heritage.

\section*{Acknowledgment}

This research is funded by Vietnam National University - Ho Chi Minh City (VNU-HCM) under Grant Number B2026-18-17.

\bibliographystyle{IEEEtran}  
\bibliography{references}   
\end{document}